\documentclass[twocolumn,showpacs,preprintnumbers,superscriptaddress,amsmath,amssymb,prl]{revtex4}

\usepackage{amssymb}
\usepackage{stmaryrd}
\usepackage{amsmath}
\usepackage{txfonts}
\usepackage{graphicx}
\usepackage{CJK}

\usepackage{bm}
\usepackage{ulem}

\begin{document}
\begin{CJK*}{GB}{SongMT}
\CJKfamily{gbsn}

\title{Generalized Poincar\'{e} Sphere}

\author{Zhi-Cheng Ren (ÈÎÖ¾³É)}
\affiliation{School of Physics and Key Laboratory of Weak-Light Nonlinear Photonics, Nankai University, Tianjin 300071, China}

\author{Yongnan Li (ÀîÓÂÄÐ)}
\affiliation{School of Physics and Key Laboratory of Weak-Light Nonlinear Photonics, Nankai University, Tianjin 300071, China}

\author{Si-Min Li}
\affiliation{School of Physics and Key Laboratory of Weak-Light Nonlinear Photonics, Nankai University, Tianjin 300071, China}

\author{Sheng-Xia Qian (Ç®Éýϼ)}
\affiliation{School of Physics and Key Laboratory of Weak-Light Nonlinear Photonics, Nankai University, Tianjin 300071, China}

\author{\\ Chenghou Tu (Í¿³Éºñ)}
\affiliation{School of Physics and Key Laboratory of Weak-Light Nonlinear Photonics, Nankai University, Tianjin 300071, China}

\author{Hui-Tian Wang (Íõ»ÛÌï)}
\email{htwang@nju.edu.cn/htwang@nankai.edu.cn}
\affiliation{School of Physics and Key Laboratory of Weak-Light Nonlinear Photonics, Nankai University, Tianjin 300071, China}
\affiliation{National Laboratory of Solid State Microstructures, Nanjing University, Nanjing 210093, China}
\affiliation{Collaborative Innovation Center of Advanced Microstructures, Nanjing 210093, China}

\date{\today}

\begin{abstract}
\noindent We present a generalized Poincar\'{e} sphere (G sphere) and generalized Stokes parameters (G parameters), as a geometric representation, which unifies the descriptors of a variety of vector fields. Unlike the standard Poincar\'{e} sphere, the radial dimension in the G sphere is not used to describe the partially polarized field. The G sphere is constructed by extending the basic Jones vector basis to the general vector basis with the \textit{continuously changeable} ellipticity (spin angular momentum, SAM) and the higher dimensional orbital angular momentum (OAM). The north and south poles of different spherical shell in the G sphere represent the pair of different orthogonal vector basis with different ellipticity (SAM) and the opposite OAM. The higher-order Poincar\'{e} spheres are just the two special spherical shells of the G sphere. We present a quite flexible scheme, which can generate all the vector fields described in the G sphere. The salient properties of this representation are able to treat the complicated geometric phase of arbitrary vector fields in the future.
\end{abstract}

\pacs{42.50.Tx, 42.25.Ja, 03.65.Vf, 42.81.--i}


\maketitle
\end{CJK*}
\newpage

Polarization is a salient nature of light, even more so than its spectral or coherence properties \cite{R01}. Polarization plays a crucial role in interaction of light with matter either linear or nonlinear. Polarization is also an important phenomenon in astronomy and has been used to study the physics of the very early universe \cite{R02}. To characterize and describe the polarization of light, there are two prominent geometric representations. The first is a widely used \textit{two-dimensional} polarization ellipse, in which three independent quantities are necessary, including the amplitudes of two orthogonal components and their phase difference, or the two principal semi-axes of the polarization ellipse and an auxiliary angle. Of course, the polarization can also be described by a pair of orthogonal right- and left-handed circularly polarized basis, $| R \rangle$ and $| L \rangle$, as

\noindent
\begin{equation} \label{01}
| A \rangle = a_R | R \rangle + a_L | L \rangle, {\text { with }} |a_R |^2 + |a_L|^2 = 1.
\end{equation}

\noindent To a certain extent, the relative refraction and phase between $a_R$ and $a_L$ can characterize the ellipticity and orientation of polarization, respectively. The second is a \textit{three-dimensional} Poincar\'{e} sphere \cite{R01}. In mathematical linguistics, the state of polarization of a \textit{polarized} light can be described by a $3 \times 1$ ``Jones vector", three normalized Stokes parameters $S_1$, $S_2$ and $S_3$, which have the following connection with $a_R$ and $a_L$

\noindent
\begin{equation}\label{02}
\left[ {\begin{array}{*{20}c}
 S_1 \\
 S_2 \\
 S_3 \\
\end{array}} \right] = \left[ {\begin{array}{*{20}c}
 \langle A | \sigma _1 | A \rangle \\
 \langle A | \sigma _2 | A \rangle \\
 \langle A | \sigma _3 | A \rangle \\
\end{array}} \right] = \left[ {\begin{array}{*{20}c}
 2\mathfrak{Re} (a_R^* a_L) \\
 2\mathfrak{Im} (a_R^* a_L) \\
 | a_R |^2 - | a_L |^2 \\
\end{array}} \right],
\end{equation}

\noindent where $\sigma _1$, $\sigma _2$ and $\sigma _3$ are the well-known Pauli matrices. $S_1$, $S_2$ and $S_3$ may be regarded as the sphere's Cartesian coordinates of a point on a unit sphere. Of course, this point may be described by the latitude $2 \theta$ and longitude $2 \varphi$, and $S_1$, $S_2$ and $S_3$ can be represented in terms of $2 \theta$ and $2 \varphi$, as follows

\noindent
\begin{equation}\label{03}
\left[ {\begin{array}{*{20}c}
 S_1 \\
 S_2 \\
 S_3 \\
\end{array}} \right] = \left[ {\begin{array}{*{20}c}
 \cos 2 \theta \cos 2 \varphi \\
 \cos 2 \theta \sin 2 \varphi  \\
 \sin 2 \theta \\
\end{array}} \right].
\end{equation}

\noindent In fact, $\varphi$ ($0 \leq \varphi < \pi$) specifies the orientation of the polarization ellipse and $\theta$ ($- \pi/4 \leq \theta \leq \pi/4$) characterizes the ellipticity and the sense of the polarization ellipse. The factor 2 in the front of $\theta$ and $\varphi$ obeys the 2$\rightarrow$1 homeomorphism between the SU(2) space of the Jones calculus and the topological SO(3) space of the standard Poincar\'{e} sphere (S sphere) $\Sigma$, and ensures that every possible state of polarization corresponds to a unique point on $\Sigma$ and vice versa. Since $\theta$ is positive (negative) according as the polarization is right-handed (left-handed), the right-handed (left-handed) polarization is represented by points on $\Sigma$ which lie the northern (southern) hemisphere of $\Sigma$, while linear polarization is represented by points on the equator. The right-handed (left-handed) circular polarization is represented by the north (south) pole.

The S sphere unifies all of the fundamental polarization descriptors. This basic geometric connection not only offers remarkable insight into but also greatly simplifies polarization problem, for instance, which has been successfully dealt with the geometric phase problem of the homogeneously polarized fields \cite{R03, R04}. When the state of polarization of a homogeneously polarized field undergoes a cyclic path on $\Sigma$, the acquired geometric phase is equal to a half of solid angle of the geodesic area subtended by the circuit on $\Sigma$. Recently, vector fields have attracted particular interest, which admit spatially-variant states of polarization \cite{R05} due to some unique features. The radially polarized vector field has the ability to produce a sharper spot beyond diffraction limit and a strong longitudinal field component by a high numerical aperture objective \cite{R06}. This has been exploited in spectroscopy \cite{R07}, particle acceleration \cite{R08}, microscopy \cite{R09}, and optical trapping \cite{R10, R11}.

In spite of its powerful utility, the S sphere is unable to deal with the vector fields. For this purpose, Milione \textit{et al}. \cite{R12} have extended it to the higher-order Poincar\'{e} sphere (H sphere), which is constructed by extending the basis of polarization in terms of the spin angular momentum (SAM) to the total optical momentum including the higher dimensional orbital angular momentum (OAM), that is to say, which is constructed by the orthogonally circular polarization basis carrying the opposite-sense OAMs. The H sphere degenerates into the S sphere in the case of zero OAM. This H sphere \cite{R12} is able to deal with the higher-order geometric phase \cite{R12, R13}. In particular, it should be emphasized that the H sphere could describe only the vector fields with the states of polarization at all the locations with the \textit{same ellipticity} but the \textit{different orientations}. Although it has been pointed out that the H sphere may be used to represent more exotic vector fields such as the hybrid vector fields \cite{R12}, it is difficult to be extended into the representation of the hybrid vector fields with their states of polarization consisting of linear, elliptical and circular polarization (i.e. spatial-variant SAM) \cite{R14, R15}. It has been demonstrated that the hybrid vector fields have novel effects and important applications \cite{R11, R16, R17}. In addition, analogous to the S sphere, Paddget \textit{et al}. \cite{R18} presented the first-order mode sphere for describing optical OAM states. The first-order mode sphere was used to explore the geometric phase associated with mode transformation of optical fields carrying OAM \cite{R19}. However, both the S sphere and the first-order mode sphere are connected to the homogenously polarized fields.

The H sphere is limited to describe the vector fields with spatially homogeneous ellipticity. A variety of vector fields, such as hybrid polarized fields, fail to appear on the H sphere. The exchange of the topological charges on the south and north poles of the H sphere leads to the completely different polarization distributions. As a result, two spheres must be used to describe the cases of the topological charge $m$~$\geq$~$1$ \cite{R12}. In particular, it is unable to conveniently describe the geometric phase between the two spheres. For the sake of breaking through the above limits, in this Letter we propose a generalized Poincar\'{e} sphere (G sphere) which extends the H sphere into more general cases. The Stokes parameters are replaced by the generalized Stokes parameters (G parameters) we newly introduced here. In the standard and higher-order Poincar\'{e} sphere representations, the radial coordinate (sphere's radius) is not used for the \textit{completely polarized} fields, while is used to describe the degree of polarization of the \textit{partially polarized} light. In contrast, in our G sphere representation, the radial coordinate is indispensable as the third freedom of degree.

Here the radial coordinate is also utilized to characterize the ellipticity of polarization, which greatly enriches the function for describing the vector fields. This dimension makes it possible to describe the general azimuthally-variant vector fields in a single spherical shell. Both the hybrid vector fields and the locally linearly polarized vector fields can be described simultaneously. We also present a scheme to generate all the vector fields described by the G sphere.

In the S sphere, in general, the north and south poles are chosen as the right- and left-handed circular polarization, respectively. In fact, no matter which pair of orthogonal basis is chosen as two poles on a S sphere, there has no essential difference after the appropriate coordinate transformation or the rotation of the sphere. Therefore, the S sphere is complete in describing all the homogeneously polarized fields, because any homogeneously polarized field can be represented by different combination of any pair of orthogonal basis. Similarly, for the first-order mode sphere, there is also no intrinsic difference whether Laguerre-Gaussian or Hermite-Gaussian modes act as the poles. So that the first-order mode sphere is also complete due to the similar reason. However, we show that the H sphere is not complete here. Once the ellipticity of the orthogonally polarized basis carrying opposite OAMs are changed, the H sphere will represent completely different high-order polarization, which cannot be recovered by rotating the sphere. To completely represent the high-order polarization, we should define the OAM-carrying orthogonal basis with continuously variable ellipticity as follows
\noindent
\begin{subequations}
\begin{align}\label{04}
|N_R^m \rangle & = \frac{1}{\sqrt{2}} e^{- j m \phi} ( e^{- j R \pi} {\bf{\hat e}}_{x} - j e^{j R \pi} {\bf{\hat e}}_{y} ), \\
|S_R^m \rangle & = \frac{1}{\sqrt{2}} e^{+j m \phi} ( e^{- j R \pi} {\bf{\hat e}}_{x} + j e^{j R \pi} {\bf{\hat e}}_{y} ).
\end{align}
\end{subequations}

\noindent where $m$ is the topological charge, $\phi$ is the azimuthal coordinate, and $R$ determines the relative phase between the $x$ and $y$ components, limited within a range of $R \in [0.5, 1]$. Clearly, $\langle S _R^m |N _R^m \rangle = 0$ is always satisfied for any $m$, implying that $|N _R^m \rangle$ and $|S _R^m \rangle$ are indeed a pair of orthogonal basis. The specific form in Eq.~(4) also satisfies the condition that all $|N _R^m \rangle$ are in phase according to the definition of Pancharatnam-phase (PB) in Ref.~\cite{R03} and all $|S _R^m \rangle$ are also in phase. This brings us the great convenience in discussion about the extended PB phase in the future. In particular, $R$ can define the ellipticity of a pair of orthogonal basis as $\epsilon_{N} = - \cos (2 R \pi) $ and $\epsilon_{S} = \cos (2 R \pi)$, which imply that both ellipticity can change continuously within a range of $[-1, 1]$ when $R \in [0.5, 1]$. Clearly, the paired basis, $|N _R^m \rangle$ and $|S _R^m \rangle$, carry the opposite SAMs and OAMs. For $R = 0.5$, the basis reduces the orthogonal circularly polarized vortices with the opposite senses (the opposite OAMs), which are the orthogonal polarization basis on the H sphere. When $m = 0$ and $R = 0.5$, the basis degenerates into the basis of the S sphere, which are the right and left circularly polarized basis.

Any azimuthally-variant vector field is composed of a combination of the basis shown in Eq.~(4)

\noindent
\begin{equation}\label{05}
| \psi^m \rangle = \psi^m_N | N^m_R \rangle + \psi^m_S | S^m_R \rangle,
\end{equation}
where $\psi^m_N$ and $\psi^m_S$ are complex coefficients indicating the fraction between two basis. If the intensity is normalized, $\psi^m_N$ and $\psi^m_S$ can be expressed as follows

\noindent
\begin{subequations}
\begin{align}\label{06}
  \psi^m_N & = \sin \beta e^{- j \phi_{0}}, \\
  \psi^m_S & =\cos \beta e^{+j \phi_{0}},
\end{align}
\end{subequations}

\noindent where $\beta \in [0, \pi/2]$ determines the fraction and $\phi_{0}$ is the additional relative phase between the two basis.

Like the Stokes parameters, the high-order Stokes parameters are extended to arbitrary orthogonal basis. Referencing the definition of the Stokes parameters in Eq.~(2) for the circularly polarized basis, we obtain the extended high-order Stokes parameters $ES_{0 R}^m$, $ES_{1 R}^m$, $ES_{2 R}^m$, and $ES_{3 R}^m$, as follows
\noindent
\begin{subequations}
\begin{align}\label{07}
 ES_{0 R}^m & = ( | \langle N_R^m | \psi^m \rangle|^{2} + | \langle S_R^m | \psi^m \rangle |^{2} ) = 1, \\
 ES_{1 R}^m & = 2 \mathfrak{Re} (\langle N_R^m | \psi^m \rangle^{*} \langle S_R^m | \psi^m \rangle ) = \sin 2 \beta \cos 2 \phi_{0}, \\
 ES_{2 R}^m & = 2 \mathfrak{Im} (\langle N_R^m | \psi^m \rangle^{*} \langle S_R^m | \psi^m \rangle ) = \sin 2 \beta \sin 2 \phi_{0}, \\
 ES_{3 R}^m & = ( | \langle N_R^m | \psi^m \rangle |^{2} - | \langle S_R^m | \psi^m \rangle |^{2} ) = -\cos 2 \beta.
\end{align}
\end{subequations}

Similarly, $| \psi^m_N |^2$ and $| \psi^m_S |^2$ are the normalized intensities of the basis, and $2 \phi_{0} = \arg(\psi^m_S)-\arg(\psi^m_N)$ is their relative phase. For the completely polarized light, the radial dimension in both the standard and H spheres is useless. Once this dimension is utilized, it will greatly enhance the sphere's ability to represent the polarization. If $R$ as the radial coordinate is introduced, the G parameters are defined as follows
\begin{subequations}
\begin{align}\label{08}
 G_{0 R}^m & =  ES_{0 R}^m R = R, \\
 G_{1 R}^m & = ES_{1 R}^m R = R \sin 2 \beta \cos 2 \phi_{0}, \\
 G_{2 R}^m & = ES_{2 R}^m R = R \sin 2 \beta \sin 2 \phi_{0}, \\
 G_{3 R}^m & = ES_{3 R}^m R = -R \cos 2 \beta.
\end{align}
\end{subequations}

The G parameters themselves are of no meaning for measurement. However, the G parameters act as a bridge between the polarization distribution of the vector field and the point on the G sphere. Referencing the definition of the Stokes parameters on the S sphere, the G sphere can be constructed by using $G_{1 R}^m$, $G_{2 R}^m$ and $G_{3 R}^m$ as the sphere's Cartesian coordinates, $G_{0 R}^m$ as the sphere's radius (the radial coordinate from the origin), and the two angles $2 \theta$ and $2 \varphi$
\begin{subequations}
\begin{align}\label{09}
R & = G_{0 R}^m, \\
\sin 2 \theta & = G_{3 R}^m/ G_{0 R}^m, \\
\tan 2 \varphi & = G_{2 R}^m/ G_{1 R}^m.
\end{align}
\end{subequations}

\noindent $2 \varphi$ and $2 \theta$ indicate the latitude and the longitude on the G sphere, as the S sphere. Compared Eq.~(9) with Eq.~(8), we obtain $2 \theta = 2 \beta-\pi/2$ and $2 \varphi = 2 \phi_{0}$. The plane determined by the axis $G_{R 1}^m$ and the sphere's axis $G_{R 3}^m$ through the north and south poles is defined as the zero meridian plane $(2 \varphi = 0)$, and the plane $2 \theta = 0$ is the equatorial plane. Any point inside the effective range $R \in [0.5, 1]$ of the G sphere can be represented by the three coordinates $(R, 2 \theta, 2 \varphi)$, which corresponds to a unique vector field with its characteristic distribution of states of polarization, like the H sphere, instead of a unique state of polarization on the S sphere. All the points on the sphere's axis (within the range of $R \in [0.5, 1]$) represent a series of scalar vortex fields with the continuously variable ellipticity. An infinite pairs of antipodal points on any spherical shell of the G sphere indicate an infinite pairs of scalar vortices with different ellipticity as orthogonal basis.
\begin{figure}[h]
  \centering{\includegraphics[width=8.5cm]{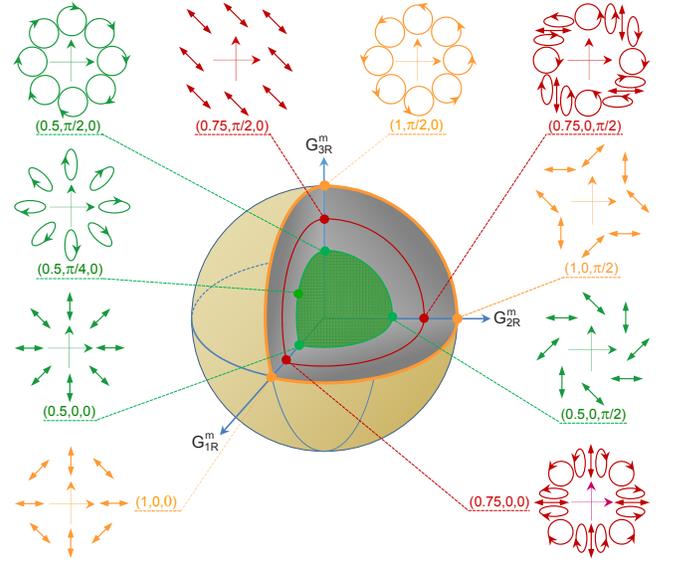}\\
  \caption{(color online) G sphere for $m = 1$. The antipodal points in the sphere's axis $G^m_{3R}$ stand for continuously varying basis. The green shell ($R = 0.5$) and the yellow shell ($R = 1$) are equivalent to the H spheres with $m = 1$ and $m = -1$, respectively. The dark red shell ($R = 0.75$) represents the vector fields with the orthogonal linearly polarized basis carrying the opposite OAMs and its equator represents the hybridly polarized vector fields.}}
\end{figure}

Let us consider a case of $m = 1$ as an example, as shown in Fig.~1. For $R = 0.5$, a pair of antipodal points $(R, 2 \theta, 2 \varphi) = (0.5, \pi/2, 0)$ and $(0.5, -\pi/2, 0)$ represent the right and left circularly polarized vortices with the topological charges of $\mp 1$, respectively. Thus any point on the spherical shell at $R = 0.5$ represents a cylindrical vector field with spatial homogeneous ellipticity. Any point at the equator of this shell indicates a local linearly polarized vector field. It is equivalent to the H sphere with $m = 1$. In contrast, on the spherical shell of $R = 1$, the SAM and the OAM of the field described by any one of the two poles have the same senses, thus this shell describes the $\pi$ vector fields, which is completely different from the case of $R = 0.5$. The spherical shell of $R = 1$ is equivalent to the H sphere with $m = -1$. The two H spheres in Ref.~\cite{R12}, just as two special shells, are successfully integrated in the G sphere.

The G sphere can also describe a variety of vector fields which cannot be described by the H sphere. On the special shell of $R = 0.75$, the basis on both poles become a pair of orthogonal linearly polarized fields carrying the opposite-sense OAMs. So this shell represents a series of non-cylindrical vector fields with azimuthally-variant ellipticity. Its equator describes all kinds of hybridly polarized vector fields with the topological charge $m$. The existence of hollow area of $R \in [0,0.5)$ is caused by the fact that the completeness of the G sphere depends only on the completeness of orthogonal basis' SAMs, regardless of principle axes of polarization ellipse. Inasmuch as the basis carrying the same SAM and OAM with different orientation of polarization are degenerate, the inner part $R \in [0,0.5)$ of the G sphere must be removed for the uniqueness of the representation.

If only $R \in [0.5, 1]$ is hold, a vector field is in one to one correspondence with a point in the G sphere. Except for the three typical spherical shells with $(R= 0.5, 0.75, 1)$, all the other shells represent more general vector fields which are orthogonal elliptic polarization as the basis. So the distribution of states of polarization for such more general vector fields exhibits the nonlinear variation of both ellipticity and orientation of local state of polarization in the azimuthal direction. Such nonlinear variation may have potential application in the future. Since the G sphere can be divided into innumerable spherical shells, the kinds of vector fields is greatly enriched compared with the H sphere. Besides, because of the completeness of the G sphere, the cases of $m = \pm 1$ are also integrated into one sphere, which greatly simplify the problem related to the vector fields, such as the high-order PB phase.
\begin{figure}[h]
  \centering{\includegraphics[width=8.5cm]{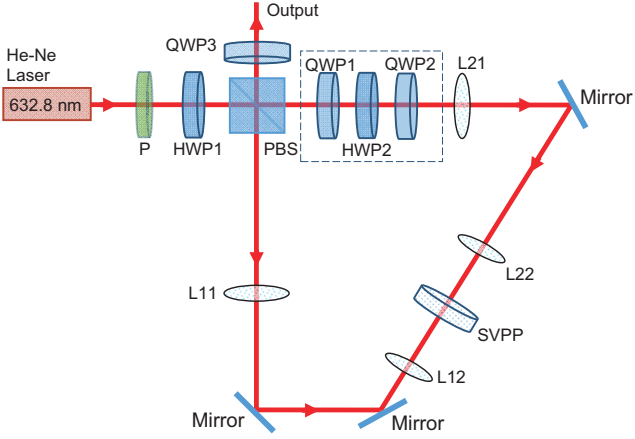}\\
  \caption{(color online) Scheme for generating all the vector fields in the G sphere.}}
\end{figure}

Based on the G sphere, as an intuitive way describing all the kinds of vector fields, a modified scheme based on the Sagnac interferometer \cite{R20} is proposed in Fig.~2, which enables flexible generation of vector fields. The polarizer P is always placed horizontally. The polarization beam splitter (PBS) divides the input field into two branches with orthogonal linear polarization and the relative intensity fraction is controlled by rotating the half wave plate (HWP1) with the angle $\alpha$ between its fast axis and the horizonal direction. Introducing the rotatable HWP1 is indispensable for generating all the vector fields in the G sphere. In the Sagnac interferometer, the spiral vortex phase plate (SVPP) with a helical phase front makes the two conterpropagating fields carry the opposite OAMs. Thus the two field owns the phase factor of $\exp(\pm j m \phi)$, respectively. Its phase function can be described by Jones matrix as
\noindent
\begin{equation}\label{10}
J_{1}= \begin{bmatrix} e^{ j m \phi} & 0 \\ 0 & e^{- j m \phi} \end{bmatrix}.
\end{equation}

Two telescopes composed of the paired lenses (L11 and L12) and (L21 and L22) image the uniform input field near in the SVPP plane to suppress the diffraction effect and to then achieve the high-quality vector fields. The HWP2 and two quarter wave plates (QWP1 and QWP2) constitute a geometric phase adjuster to flexibly control the relative phase between two branches, where the fast axes of the QWP1 and QWP2 are fixed at $+45^\circ$ with the horizonal direction and the HWP2 is rotatable with its fast axis forming an angle $\theta_{2}$ with the horizonal direction. Thus this geometric phase adjuster can be expressed by Jones matrix as
\noindent
\begin{equation}\label{11}
J_{2}= \begin{bmatrix} e^ {- j (2\theta_{2} + \pi/2)} & 0 \\ 0 & e^ { j (2\theta_{2} + \pi/2)} \end{bmatrix}.
\end{equation}

The rotatable QWP3, with its fast axis forming an angle $\theta_{3}$ with the horizonal direction, is expressed by Jones matrix as
\noindent
\begin{equation}\label{12}
J_{3}= \frac{1}{2}\begin{bmatrix} 1 - j \cos (2\theta_{3}) & - j \sin (2\theta_{3}) \\ - j \sin (2\theta_{3}) & 1+ j \cos (2\theta_{3}) \end{bmatrix}.
\end{equation}

We define the input field before the PBS as $\mathbf{\hat{P}}_{in}$ and the output field behind the QWP3 as $\mathbf{\hat{P}}_{out}$. The normalized $\mathbf{\hat{P}}_{in}$ and $\mathbf{\hat{P}}_{out}$ can be expressed as
\noindent
\begin{align}\label{13,14}
\mathbf{\hat{P}}_{in} & = \cos (2 \alpha) \mathbf{\hat{e}}_{x} + \sin (2 \alpha) \mathbf{\hat{e}}_{y}, \\
\mathbf{\hat{P}}_{out} & = J_{3}J_{2}J_{1} \mathbf{\hat{P}}_{in} = \sin (2\alpha) \exp [ + j (2\theta_{2} + \pi /2)] \mathbf{\hat{e}}_{N} \nonumber \\
 & \quad \quad \quad \quad \quad + \cos (2\alpha) \exp [ - j (2\theta_{2} + \pi /2)] \mathbf{\hat{e}}_{S},
\end{align}
\noindent where
\noindent
\begin{subequations}
\begin{align}\label{15}
\mathbf{\hat{e}}_{N} & = \frac{1}{\sqrt{2}} e^{- j m \phi} \left\{[1 + j \cos (2\theta_{3}) ] \mathbf{\hat{e}}_{y} - j \sin (2\theta_{3}) \mathbf{\hat{e}}_{x} \right\}, \\
\mathbf{\hat{e}}_{S} & = \frac{1}{\sqrt{2}} e^{+ j m \phi} \left\{ [1 - j \cos (2\theta_{3})] \mathbf{\hat{e}}_{x} - j \sin (2\theta_{3}) \mathbf{\hat{e}}_{y} \right\}.
\end{align}
\end{subequations}

Compared Eq.~(15) with Eq.~(4), both share similarity. The paired orthogonally polarized basis carry the opposite OAMs, and both SAMs can take all the allowed value because $R$ in Eq.~(4) or $\theta_{3}$ in Eq.~(15) can be changed continuously. However, both have a little difference that the orientations of the principle axes of the polarization in Eq.~(4) keep invariant, while they are changeable in Eq.~(15). The output field has no relation to the orientations of the basis in despite of a relative rotation about the sphere's axis of the G sphere. It needs no mind because which can be compensated by adjusting the relative phase by rotating the HWP2. The key factor dominating the output field is the SAM (ellipticity) of the orthogonal basis, which is controlled by the QWP3. Compared Eq.~(14) with Eq.~(6), we have $\beta = 2 \alpha$ and $\phi_{0} = - (2 \theta_{2} + \pi/2)$. Clearly, the HWP1 controls the latitude, the HWP2 the longitude, and the QWP3 the radius of the shell in the G sphere.

To demonstrate the reliability of this scheme, we perform the experiment by a continuous-wave linearly polarized He-Ne laser as the source and a SVPP, operating at a wavelength of 632.8~nm. For simplicity and without loss of generality, only the case of $m = 1$ is performed. Figure~3 shows the generated vector fields which are five among the ten fields shown in Fig.~1, with the spherical coordinates $(R, 2\theta, 2\varphi) = (0.5, 0, 0)$, $(0.5, \pi/4, 0)$, $(0.75, 0, 0)$, $(1, 0, 0)$ and $(0.5, \pi/2, 0)$. As shown in Fig.~1, these five points on the G sphere represent the radially polarized vector field, the radially elliptically polarized vector field, the hybridly polarized vector field, the $\pi$-vector field, and the homogeneously right circularly polarized vortex field, respectively. The third and fourth rows in Fig.~3 show the intensity patterns behind the horizonal and vertical polarizers, respectively, which are in good agreement with the theoretical prediction.

\begin{figure}[h]
  \centering{\includegraphics[width=8.5cm]{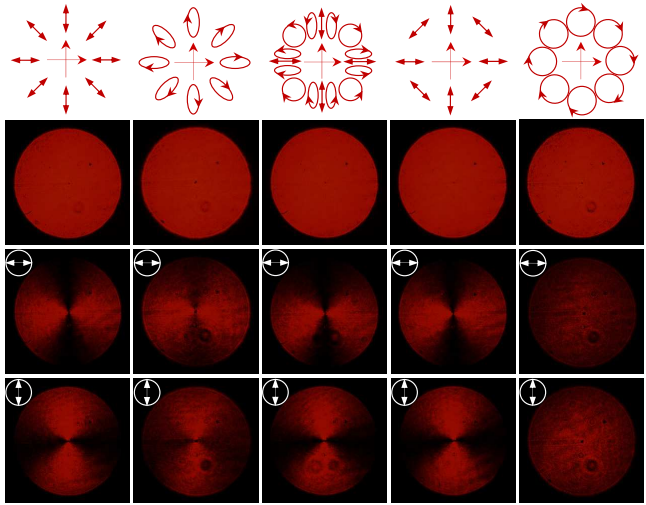}\\
  \caption{(color online) Experimentally generated vector fields described in the G sphere. First row shows the distributions of states of polarization. Second row shows the corresponding vector fields. Third and fourth rows indicate the intensity patterns behind a horizonal and vertical polarizers, respectively.}}
\end{figure}

In summary, we have constructed the G sphere, which effectively utilizes the inner part of the S sphere to unify the descriptors of a variety of vector fields. This G sphere, as a geometric connection, possesses many advantages in describing the vector fields, and provides not only remarkable insight into but also greatly simplifies otherwise complex polarization problems, and as a result has become an ubiquitous device with which to treat polarization phenomena in numerous and varying fields such as the general high-order PB phase, in particular, high-order PB phase problem between the vector fields with different topological charges. Besides, based on the G sphere we propose a reliable and convenient scheme to generate all the vector fields expressed by this sphere.

This work is supported by the 973 Program of China under Grant No. 2012CB921900 and the National scientific instrument and equipment development project 2012YQ17004, the National Natural Science Foundation of China under Grants 11274183 and 11374166, Tianjin research program of application foundation and advanced technology 13JCZDJC33800 and 12JCYBJC10700.


\end{document}